\documentclass[a4paper,11pt]{article}
\usepackage{pos}
\usepackage{graphicx}
\usepackage{subcaption}
\usepackage{xspace}
\usepackage{amsmath}
\usepackage{booktabs}
\usepackage{url}
\graphicspath{{proceeding_figures/}}

\title{High-$p_{\rm T}$ physics and jet production}
\ShortTitle{High-$p_{\rm T}$ physics and jet production}

\author*[a,b]{Francesco Giuli}

\affiliation[a]{Universit\`a degli Studi Link,  Via del Casale di S. Pio V, 44, 00165\
Rome, Italy}

\affiliation[b]{INFN Sezione di Roma Tor Vergata, Via della Ricerca Scientifica, 1, 00133,\
Rome, Italy}

\emailAdd{francesco.giuli@cern.ch}

\abstract{Jet production is the dominant high-$p_{\rm T}$ process at hadron colliders and provides a central testing ground for perturbative QCD, parton distribution functions and determinations of the strong coupling.  This contribution summarises recent measurements of inclusive-jet, dijet and jet-multiplicity observables presented at DIS2026, with emphasis on the interplay between experimental precision, next-to-next-to-leading-order predictions, non-perturbative and electroweak corrections, and the treatment of correlated systematic uncertainties.  Inclusive jet measurements from CMS and ATLAS constrain the gluon distribution at large Bjorken $x$ and enable extractions of $\alpha_s(m_Z)$ compatible with the world average.  Dijet measurements provide complementary sensitivity through the dijet invariant mass, rapidity separation and longitudinal boost, while ratios of inclusive jet multiplicities reduce several experimental and PDF uncertainties and directly probe additional QCD radiation.  Recent progress in jet-energy calibration, together with new results from RHIC, ALICE and CMS on jet substructure and heavy-quark radiation, illustrates the breadth of current high-$p_{\rm T}$ jet physics and its relevance for Run~3 and future global PDF analyses.}

\FullConference{The 33rd International Workshop on Deep Inelastic Scattering and Related Subjects (DIS2026)\\
4--8 May 2026\\
Bologna, Italy\\}

\begin{document}
\maketitle

\section{Introduction}

Jets are the most abundant high-transverse-momentum final state at hadron colliders.  Their production rates span many orders of magnitude in cross section and probe hard scales from a few tens of GeV to several TeV.  As a consequence, jet measurements simultaneously test the perturbative expansion of QCD, the modelling of parton showers and hadronisation, and the proton structure encoded in parton distribution functions (PDFs).  In global fits, inclusive-jet and dijet data are among the most direct constraints on the gluon PDF at medium and large Bjorken $x$, and they are strongly correlated with determinations of the strong coupling $\alpha_s$.

The recent experimental programme has moved from precision measurements of single spectra to multi-differential observables with full covariance information.  The theory description has also improved substantially: next-to-next-to-leading-order (NNLO) QCD predictions are now available for several inclusive-jet and dijet measurements, often through interpolation grids, and are supplemented by non-perturbative (NP) and electroweak (EW) corrections.  The resulting comparisons are precise enough that the dominant limitations are frequently the jet energy scale (JES), the modelling of systematic uncertainty correlations, the residual missing-higher-order uncertainty, and the mutual compatibility of different PDF data sets.

This proceeding summarises selected results from the DIS2026 plenary talk on high-$p_{\rm T}$ physics and jet production.  The focus is on recent LHC jet measurements and their impact on PDFs and $\alpha_s$, followed by examples extending the same methods to RHIC energies, heavy-flavour radiation, boosted electroweak objects and searches.

\section{Inclusive jet production and PDF constraints}

The CMS measurement of double-differential inclusive jet cross sections at $\sqrt{s}=13$~TeV uses anti-$k_t$ jets with $R=0.4$ and $R=0.7$ and is measured as a function of jet $p_{\rm T}$ and rapidity~\cite{CMS:InclusiveJets13TeV}.  In the main kinematic region the experimental uncertainty is below about $5\%$, with the JES giving the dominant contribution.  Comparisons to fixed-order QCD show the importance of NNLO corrections: the data are better described than at NLO and the scale uncertainty is reduced.  NP corrections are relevant at low scales, where generator differences dominate their uncertainty, while EW corrections become important for central, high-energy jets.  The same measurement enables a simultaneous PDF and $\alpha_s$ extraction, giving $\alpha_s(m_Z)=0.1166\pm0.0017$ in the addendum to Ref.~\cite{CMS:InclusiveJets13TeV}.

\begin{figure}[t!]
  \centering
  \includegraphics[width=0.82\linewidth]{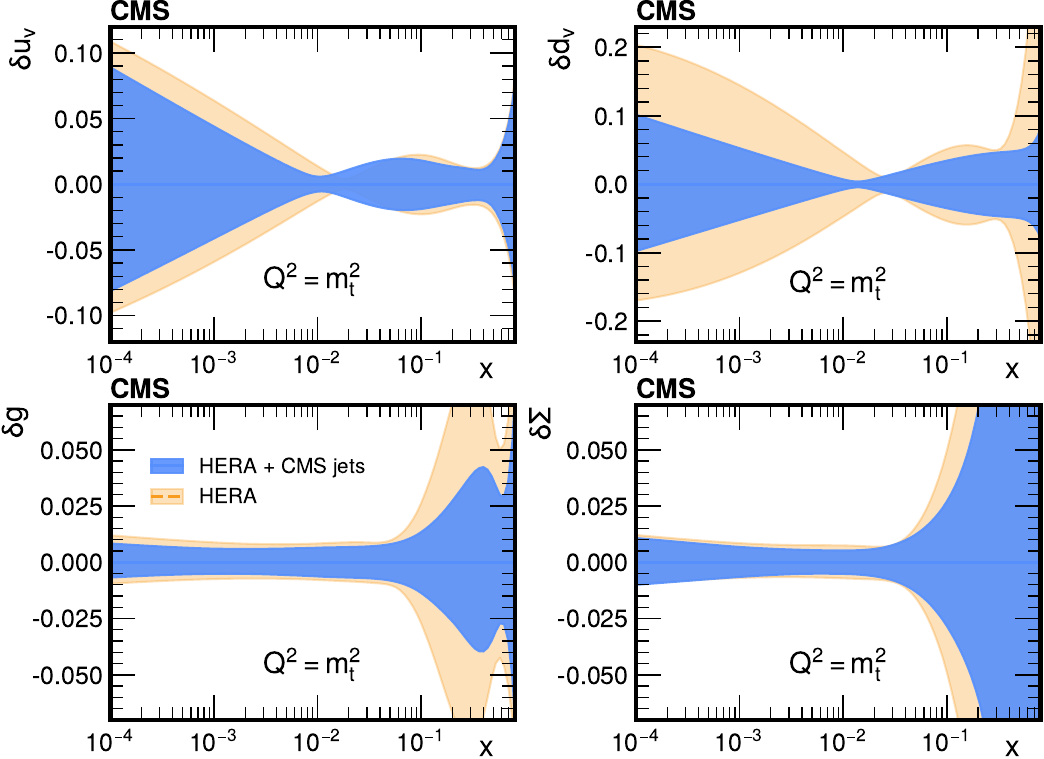}
  \caption{Impact of the combined CMS inclusive-jet data at four centre-of-mass energies on the valence-quark, gluon and sea-quark PDF uncertainties.  Plots taken from Ref.~\cite{CMS:InclusiveJetsCombination}.}
  \label{fig:cms-inclusive}
\end{figure}

A further CMS analysis combines inclusive-jet measurements at four centre-of-mass energies, $\sqrt{s}=2.76$, 7, 8 and 13~TeV, with HERA inclusive DIS data~\cite{CMS:InclusiveJetsCombination,HERA:HERAPDF20}.  The multi-energy combination extends the lever arm in $x$ and scale, reduces uncertainties in the down-valence and gluon PDFs compared with HERA-only fits, and extracts a value of $\alpha_s$ compatible with the Particle Data Group average at NNLO.  Fig.~\ref{fig:cms-inclusive} illustrates the resulting $\alpha_s$ comparison and PDF impact.

\begin{figure}[t]
  \centering
  \includegraphics[width=0.43\linewidth]{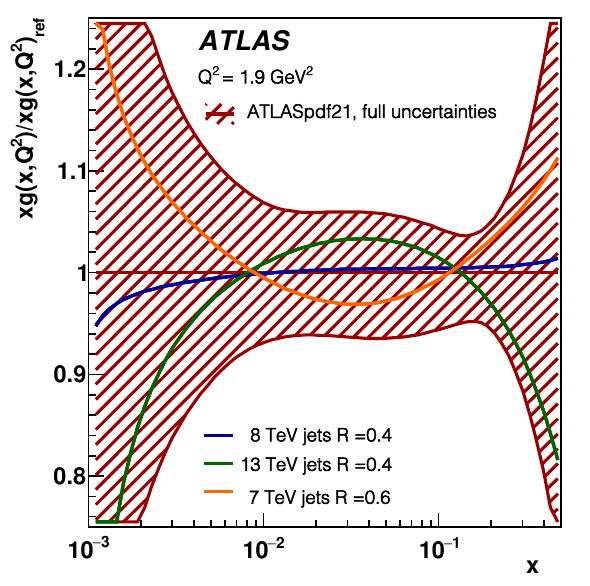}
  \includegraphics[width=0.43\linewidth]{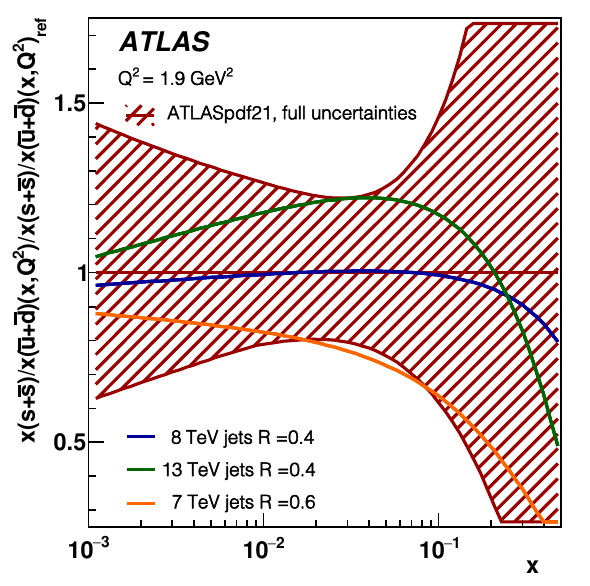}
  \caption{Effect of various different jet production data sets at 7, 8 and 13 TeV, with differing choices of jet radius, on the gluon PDF and on the strange-to-light sea ratio. Uncertainties of the central fit are full uncertainties: experimental, evaluated with tolerance T=3, plus model and parameterisation uncertainties.  Plots taken from Ref.~\cite{ATLAS:ATLASpdf21}.}
  \label{fig:atlaspdf}
\end{figure}

ATLAS has performed a complementary global fit, ATLASpdf21, using a broad set of ATLAS measurements together with HERA DIS data~\cite{ATLAS:ATLASpdf21}.  The HERA data provide the backbone over a wide range of $x$ and $Q^2$, while LHC measurements add constraints at medium and high $x$ and at high scales.  The ATLASpdf21 fit is performed with the xFitter framework~\cite{xFitter} and includes NNLO QCD and NLO EW effects either through direct NNLO grids or through $K$-factor corrections.  Inclusive-jet data mildly harden the high-$x$ gluon and considerably reduce its uncertainty.  This is visible in Fig.~\ref{fig:atlaspdf}: compared with the no-jet fit, the inclusive-jet input narrows the high-$x$ gluon band while leaving the strange-to-light sea ratio largely stable within the quoted uncertainties.  Fits using 7, 8 or 13~TeV inclusive-jet inputs show differences mostly driven by the centre-of-mass energy, but these differences are not significant compared with the full PDF uncertainties when the ATLAS tolerance prescription is applied.

The treatment of correlated systematic uncertainties is a recurring issue in these comparisons.  For example, in the ATLAS 8~TeV inclusive-jet measurement~\cite{ATLAS:InclusiveJets8TeV}, alternative decorrelation prescriptions for jet-flavour response and other components affect the global $\chi^2$ more strongly than the fitted PDFs.  This highlights that a precise covariance model is as important as the central measurement when jet data are used in global fits.

\section{Dijet production}

Dijet observables provide an independent handle on the hard scale and on the incoming parton momentum fractions.  CMS measured double- and triple-differential dijet cross sections at $\sqrt{s}=13$~TeV using anti-$k_t$ jets with $R=0.4$ and $R=0.8$~\cite{CMS:Dijet13TeV}.  The two-dimensional measurement is performed as a function of the dijet invariant mass $m_{1,2}$ and the maximum rapidity of the two leading jets, while the three-dimensional measurement uses additional rapidity variables.  The rapidity separation is particularly sensitive to new high-mass dynamics, whereas the longitudinal boost probes the PDF dependence.

\begin{figure}[t]
  \centering
  \includegraphics[width=0.9\linewidth]{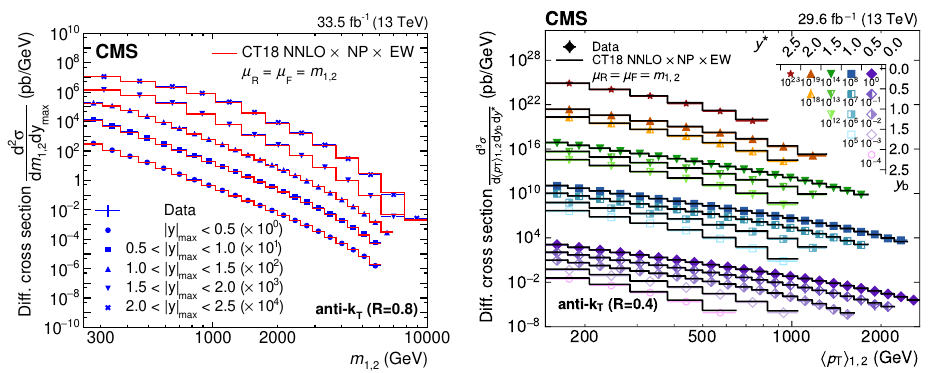}
  \caption{Differential dijet cross sections, illustrated here for the (left) 2D measurement as a function of $m_{1,2}$ using jets with R = 0.8, and the (right) 3D measurement as a function of $<p_{\mathrm{T}}>_{1,2}$ using jets with R= 0.4. The markers and lines indicate the measured unfolded cross sections and the corresponding NNLO predictions, respectively. Plots taken from Ref.~\cite{CMS:Dijet13TeV}.}
  \label{fig:cms-dijet}
\end{figure}

The CMS dijet predictions are computed at NNLO with \textsc{NNLOJET} and fastNLO interpolation grids, and are multiplied by NP and EW corrections.  The scale uncertainty is greatly reduced at NNLO.  The data are generally well described, although different PDF sets differ at high $p_{\rm T}$ or high invariant mass; in the comparison shown in the talk, MSHT20 provides one of the best descriptions while ABMP16 gives a poorer one.  Fig.~\ref{fig:cms-dijet} shows representative double- and triple-differential spectra: the steeply falling $m_{1,2}$ distribution demonstrates the large dynamic range, while the different rapidity slices separate regions with different typical Bjorken-$x$ values.  The QCD analysis of the 13~TeV dijet data with HERA DIS data yields $\alpha_s(m_Z)=0.1179$ for the 2D analysis and $0.1181$ for the 3D analysis, with total uncertainties of order $0.002$, and constrains the gluon PDF mainly for $x\gtrsim0.1$~\cite{CMS:Dijet13TeV}.

\begin{figure}[t]
  \centering
  \includegraphics[width=0.43\linewidth]{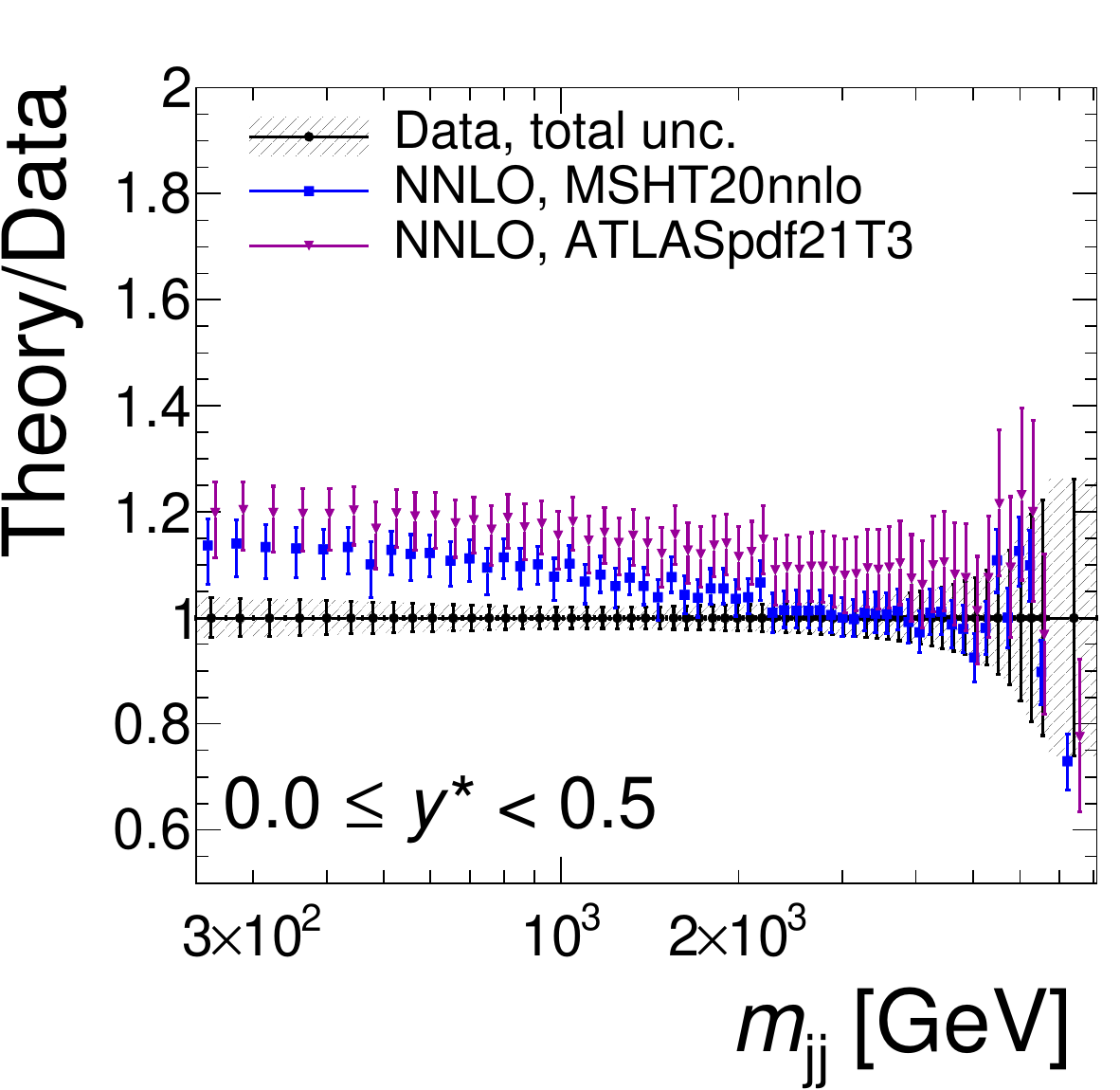}
    \includegraphics[width=0.43\linewidth]{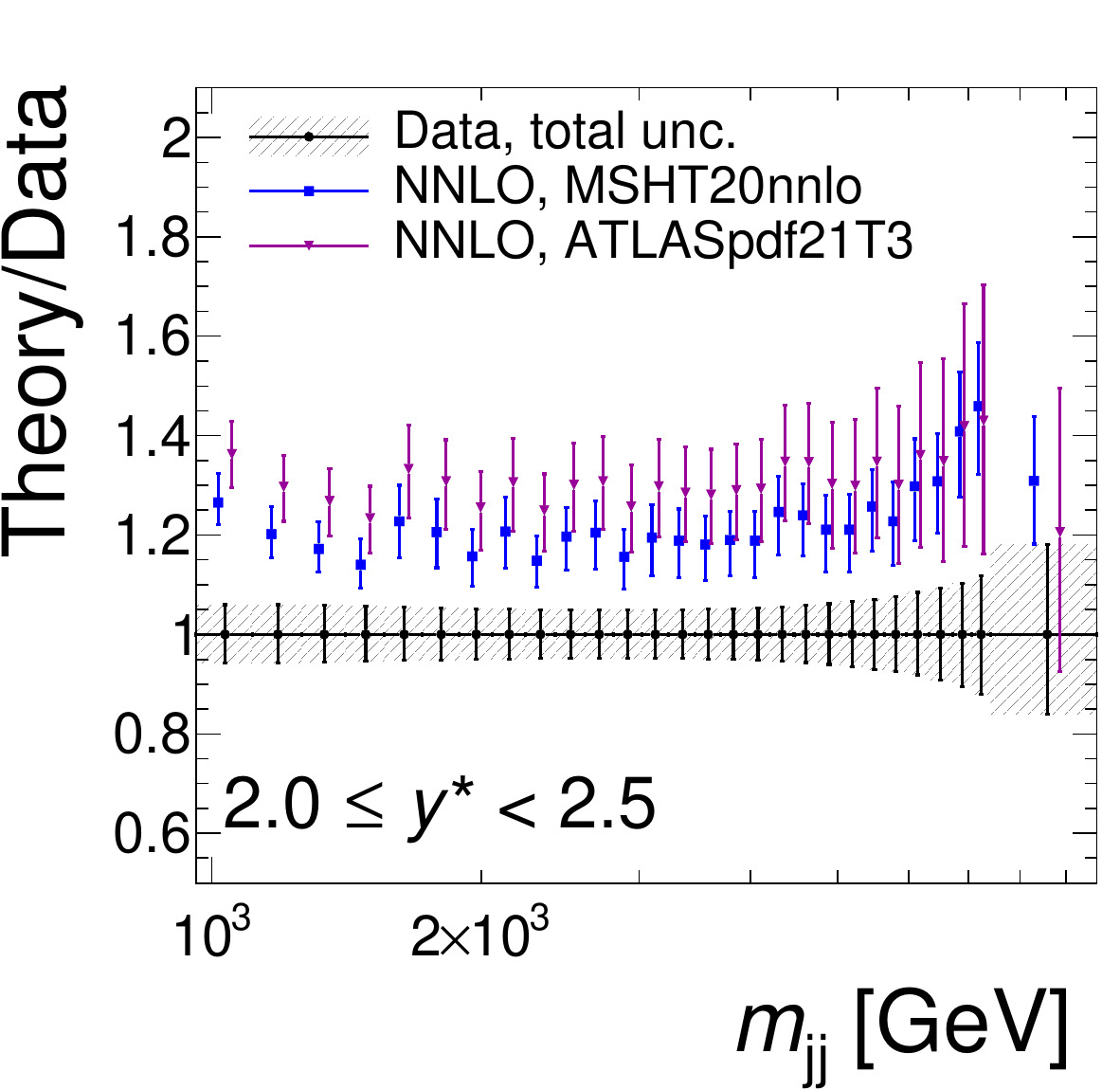}
  \caption{ATLAS full-Run~2 dijet measurement: Theory-to-data ratio for the double-differential cross-section as a function of the invariant dijet mass $m_{jj}$ in 2 representative $y^{*}$ bins for anti-k$_{\mathrm{t}}$ R=0.4 jets at the particle level. The sub-figures compare NNLO pQCD calculations, corrected by electroweak and non-perturbative corrections, with different PDF sets (MSHT, ATLASpdf21T3) to the ATLAS data. Coloured vertical line errors indicate the total systematic uncertainties in pQCD calculations. Vertical band errors around 1.0 indicate total experimental uncertainty in the data. Plots taken from Ref.~\cite{ATLAS:DijetFullRun2}.}
  \label{fig:atlas-dijet}
\end{figure}

ATLAS has recently extended its dijet programme to the full Run~2 data set, measuring inclusive dijet cross sections at $\sqrt{s}=13$~TeV with 140~fb$^{-1}$~\cite{ATLAS:DijetFullRun2}.  The measurement supersedes the earlier 3.2~fb$^{-1}$ result~\cite{ATLAS:JetsDijets13TeV}.  It covers dijet masses from about 240~GeV to almost 10~TeV and is presented both in $(m_{jj},y^*)$ and $(m_{jj},y_{\rm boost})$ bins, with full statistical and systematic covariance information.  Fig.~\ref{fig:atlas-dijet} shows that the central theory predictions for several modern PDF sets differ at a level comparable to the experimental precision.  Experimental and theoretical uncertainties are both at the few-percent level, making the comparison very sensitive to long-range correlations and normalisation shifts.  Individual rapidity bins often show reasonable agreement with NNLO predictions, while global $\chi^2$ values can be large; the best quoted agreement is obtained for ATLASpdf21 with tolerance $T=3$, partly because of its larger PDF uncertainty.

\section{Jet multiplicity ratios and the role of calibration}

Ratios of inclusive jet multiplicities provide a different route to precision QCD.  ATLAS measured ratios such as
\begin{equation}
R_{3/2}=\frac{d\sigma_{\geq 3\,\mathrm{jet}}/dx}{d\sigma_{\geq 2\,\mathrm{jet}}/dx},\quad
R_{4/2}=\frac{d\sigma_{\geq 4\,\mathrm{jet}}/dx}{d\sigma_{\geq 2\,\mathrm{jet}}/dx},\quad
R_{4/3}=\frac{d\sigma_{\geq 4\,\mathrm{jet}}/dx}{d\sigma_{\geq 3\,\mathrm{jet}}/dx},
\end{equation}
and analogous higher-multiplicity ratios in 140~fb$^{-1}$ of 13~TeV data~\cite{ATLAS:JetRatios}.  Many NP and PDF effects cancel in these ratios, increasing their sensitivity to $\alpha_s$ and to the modelling of extra QCD radiation.  The observables include $H_{T,2}=p_{T,1}+p_{T,2}$, which tests perturbative convergence, and $\Delta y_{jj}$ or $m_{jj}$, which are relevant to vector-boson-fusion and vector-boson-scattering modelling.

\begin{figure}[t]
  \centering
  \includegraphics[width=0.43\linewidth]{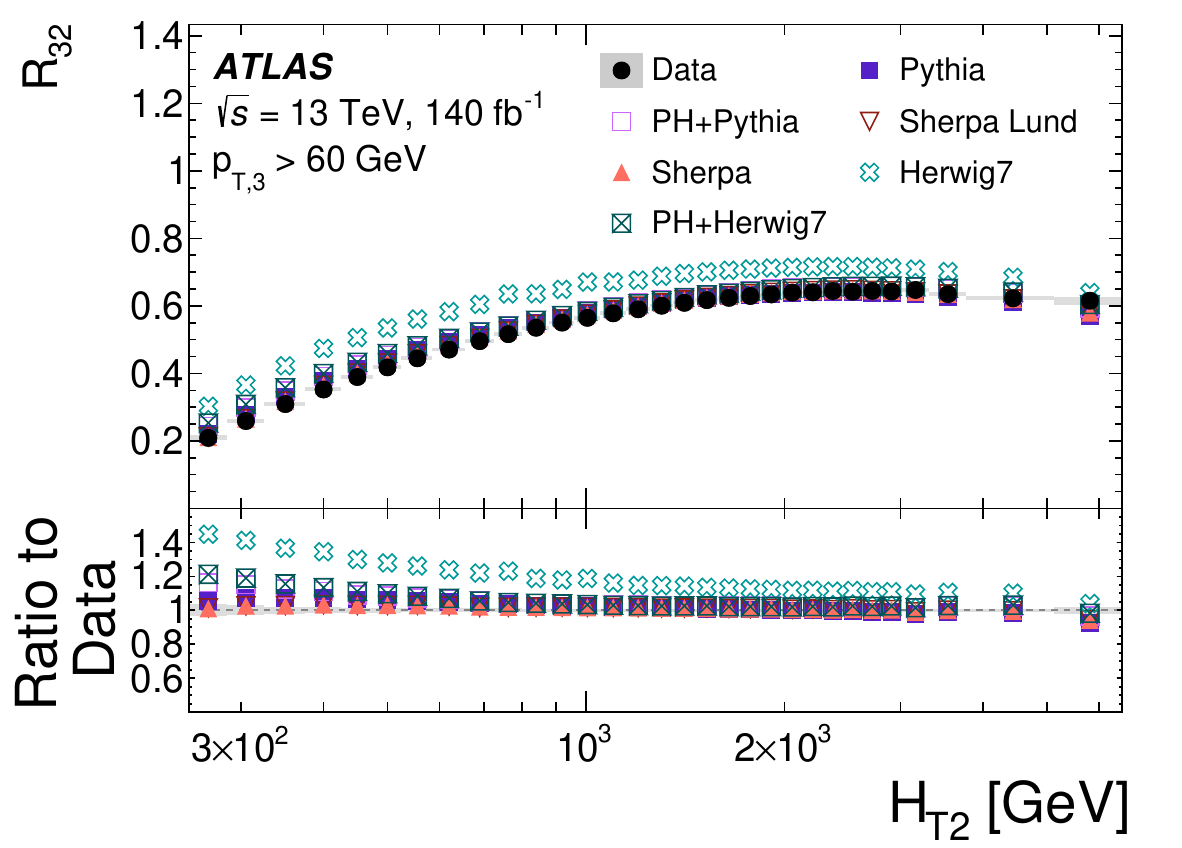}
  \includegraphics[width=0.43\linewidth]{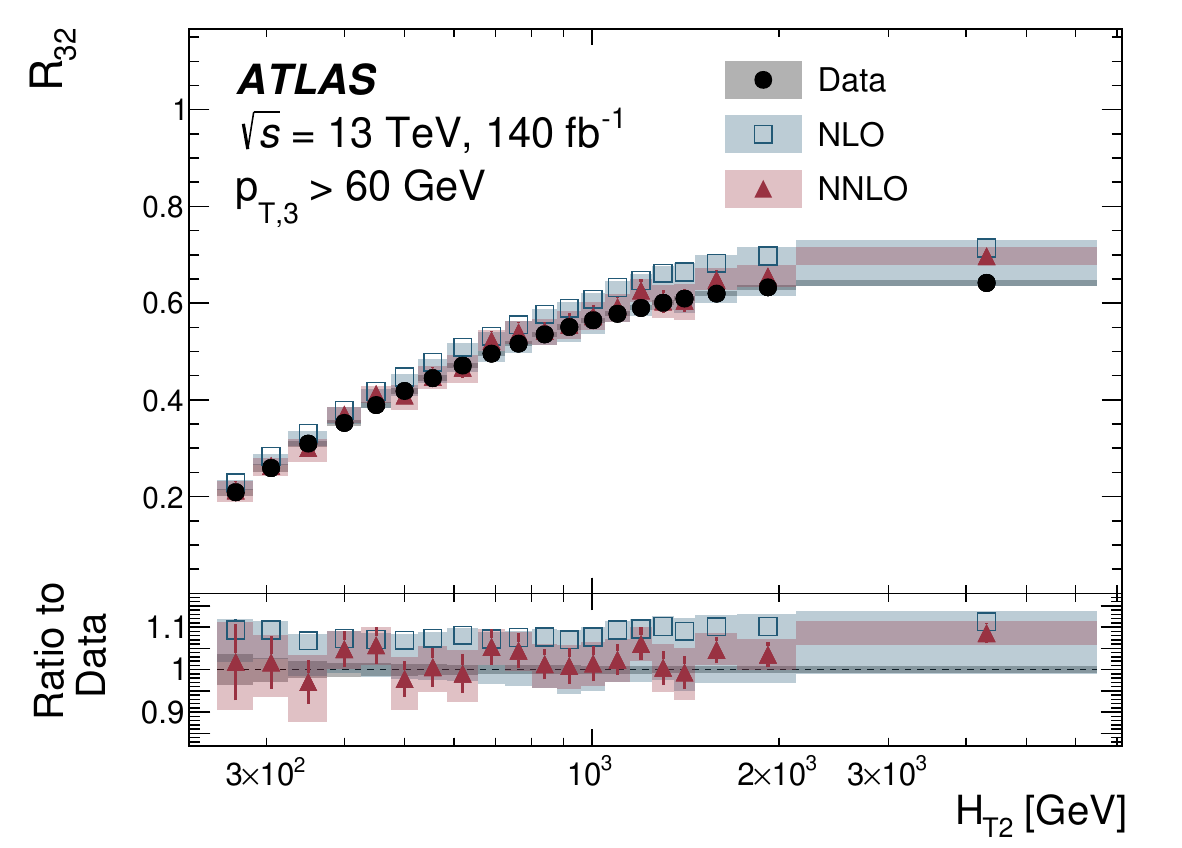}
  \caption{ATLAS jet-multiplicity ratios $R_{32}$ as functions of $H_{\mathrm{T},2}$ for $p_{\mathrm{T},3}$. The data error bands show the statistical and systematic components summed in quadrature. The theory error bands include contributions from the statistical, PDF, and scale variations. The statistical uncertainty on the theory predictions is illustrated with a vertical line. The lower figure panels provide ratios of the predictions to the unfolded data. Plots taken from Ref.~\cite{ATLAS:JetRatios}.}
  \label{fig:ratios}
\end{figure}

NNLO predictions describe both the value and shape of $R_{3/2}$ more accurately than NLO predictions, which tend to overestimate the data.  The improvement is reflected in smaller $\chi^2$/ndf values over a broad range of third-jet $p_{\rm T}$ thresholds.  Fig.~\ref{fig:ratios} illustrates the behaviour for two threshold definitions: the ratio increases with $H_{T,2}$ when the probability of an additional hard jet grows, while a more restrictive third-jet requirement suppresses the inclusive three-jet rate.  Comparisons to parton-shower generators reveal additional modelling challenges at higher multiplicities and in large-$m_{jj}$ regions, where logarithms of the hard-scattering energy over jet transverse momentum can become important.

\begin{figure}[t!]
  \centering
    \includegraphics[width=0.45\linewidth]{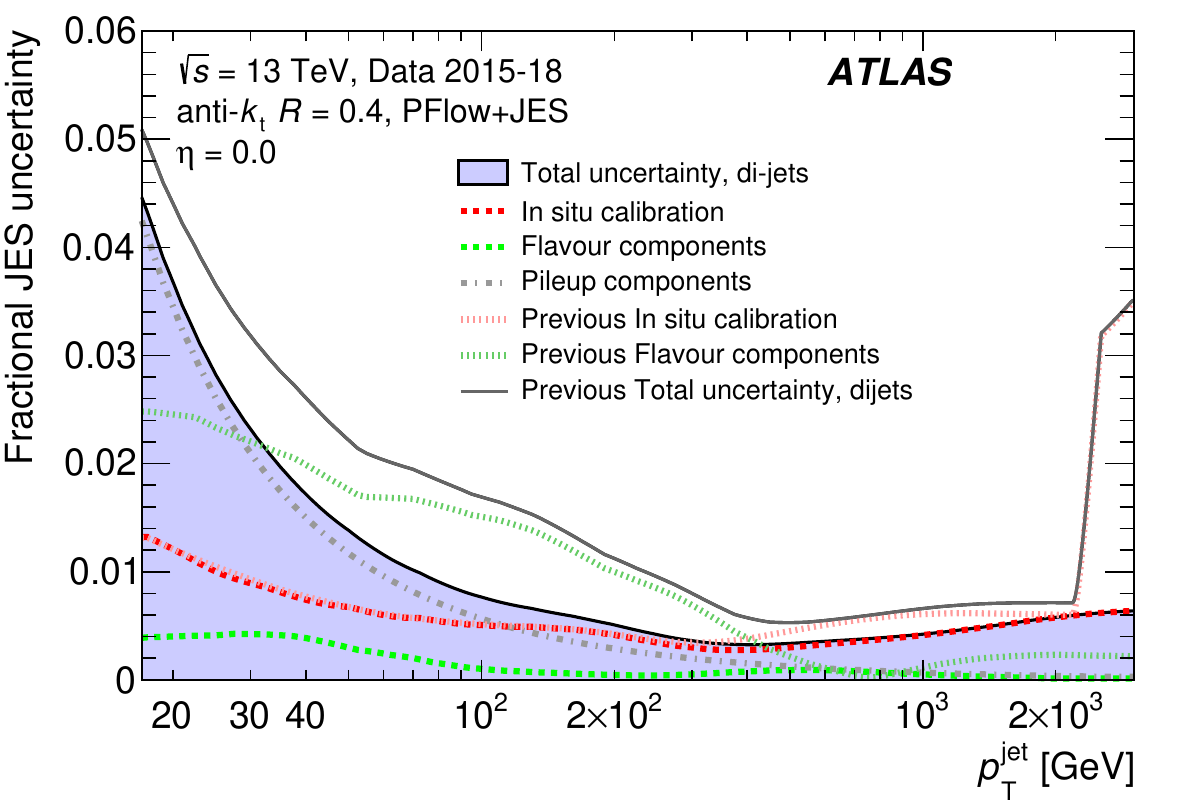}
    \includegraphics[width=0.48\linewidth]{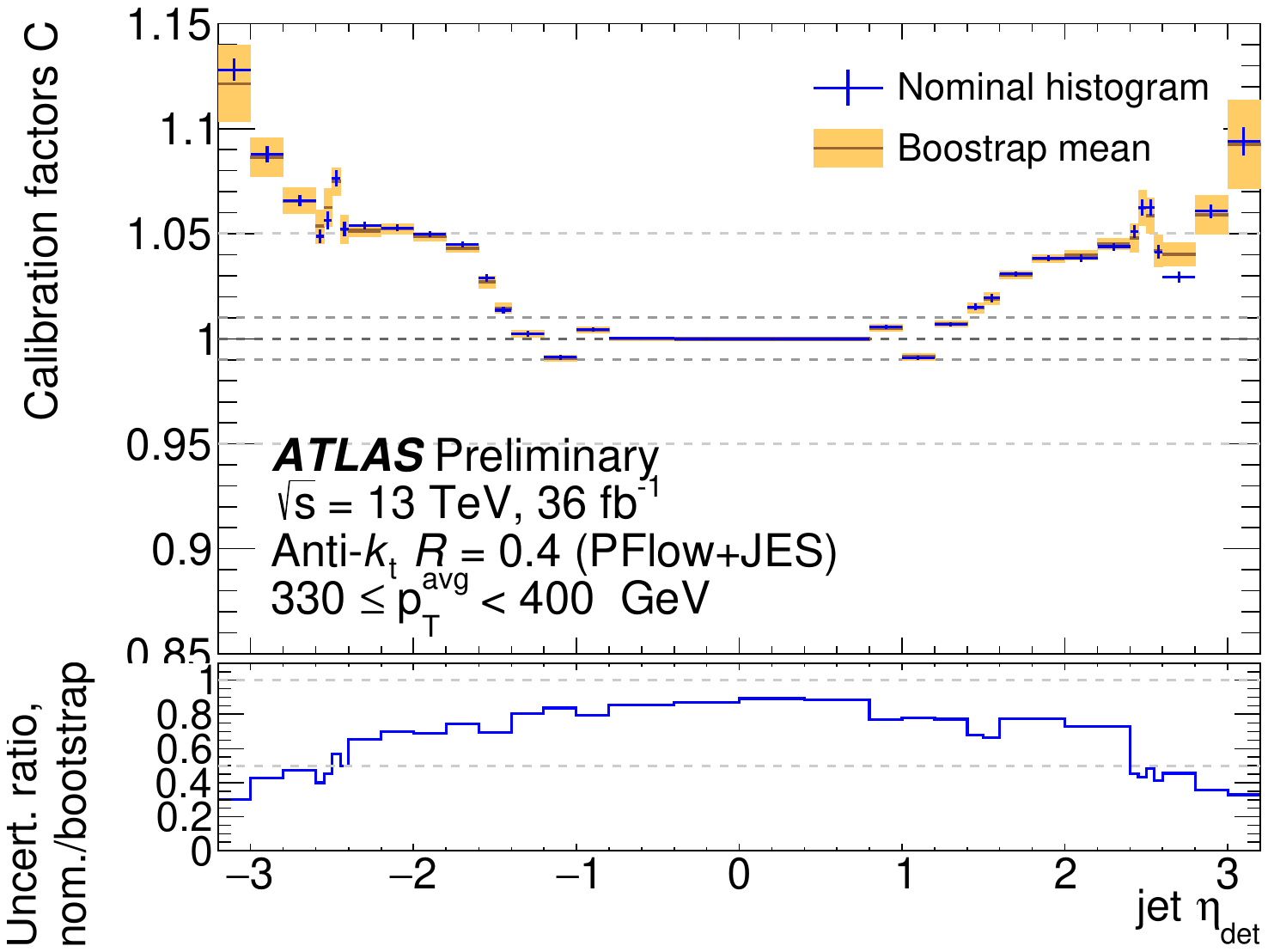}
  \caption{(Left) Total fractional JES for PFlow jets in the most central pseudo-rapidity region as a function of the jet $p_{\mathrm{T}}$. The blue area shows the quadrature sum of all the uncertainties. For comparison the uncertainties as previously recommended for physics analysis are also given. Plot taken from Ref.~\cite{ATLAS:JESsingleParticle}. (Right) Calibration factors for various $p_{\mathrm{T}}^{\mathrm{avg}}$ ranges. The statistical uncertainty propagated with the previous approach is in dark blue, while the statistical uncertainty propagated with the bootstrap method is in yellow. The bottom panel displays the ratio of the statistical uncertainties evaluated with the two approaches. Plot taken from Ref.~\cite{ATLAS:BootstrapJES}.}
  \label{fig:jes}
\end{figure}

The precision of these measurements relies on recent improvements in the JES.  ATLAS has derived jet-energy corrections and uncertainties from single-particle response measurements in data and simulation, extending the kinematic reach by using isolated pions from $\tau$ decays~\cite{ATLAS:JESsingleParticle}.  Compared with the traditional $p_{\rm T}$-balance method, the approach gives a roughly 20\% improvement for $300<p_{\rm T}<2000$~GeV and a much larger improvement in the very-high-$p_{\rm T}$ region where the balance method has limited reach.  The combined JES uncertainty reaches about 0.3\% at 300~GeV and 0.6\% at 1~TeV in the central region.  Fig.~\ref{fig:jes} (left) shows this reduction directly: the total uncertainty for central jets is now at the sub-percent level over the most relevant high-$p_{\rm T}$ range, so the calibration is precise enough that correlation modelling becomes a leading issue in PDF applications.

A second, closely related, ingredient is the statistical treatment of the JES $\eta$-intercalibration.  The $\eta$-intercalibration propagates the well-constrained central-jet response to the full detector acceptance, and its statistical uncertainty can induce long-range correlations across rapidity and $p_{\rm T}$.  Ref.~\cite{ATLAS:BootstrapJES} introduces a bootstrap approach to derive these uncertainties and correlations directly from resampled calibration inputs.  In Fig.~\ref{fig:jes} (right), the nominal calibration factors are compared with the bootstrap mean in representative $p_{\rm T}^{\rm avg}$ intervals; the lower panels show that the uncertainty estimated by the previous nominal method can be smaller than the bootstrap estimate in sizeable regions.  This matters for high-precision jet cross sections because overly rigid or underestimated intercalibration correlations can artificially increase the global $\chi^2$ in PDF fits, or conversely hide tensions that are due to detector modelling rather than to QCD.

\begin{figure}[t!]
  \centering
    \includegraphics[width=0.9\linewidth]{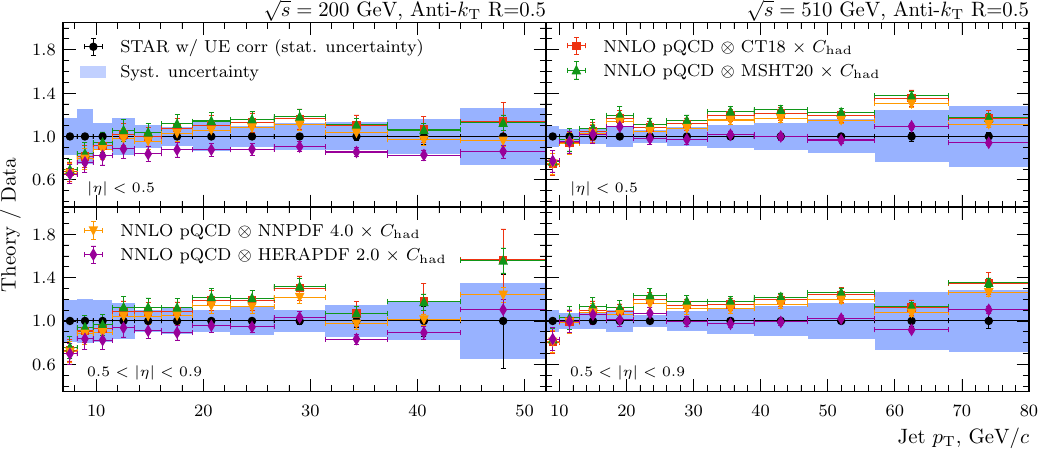}
    \includegraphics[width=0.52\linewidth]{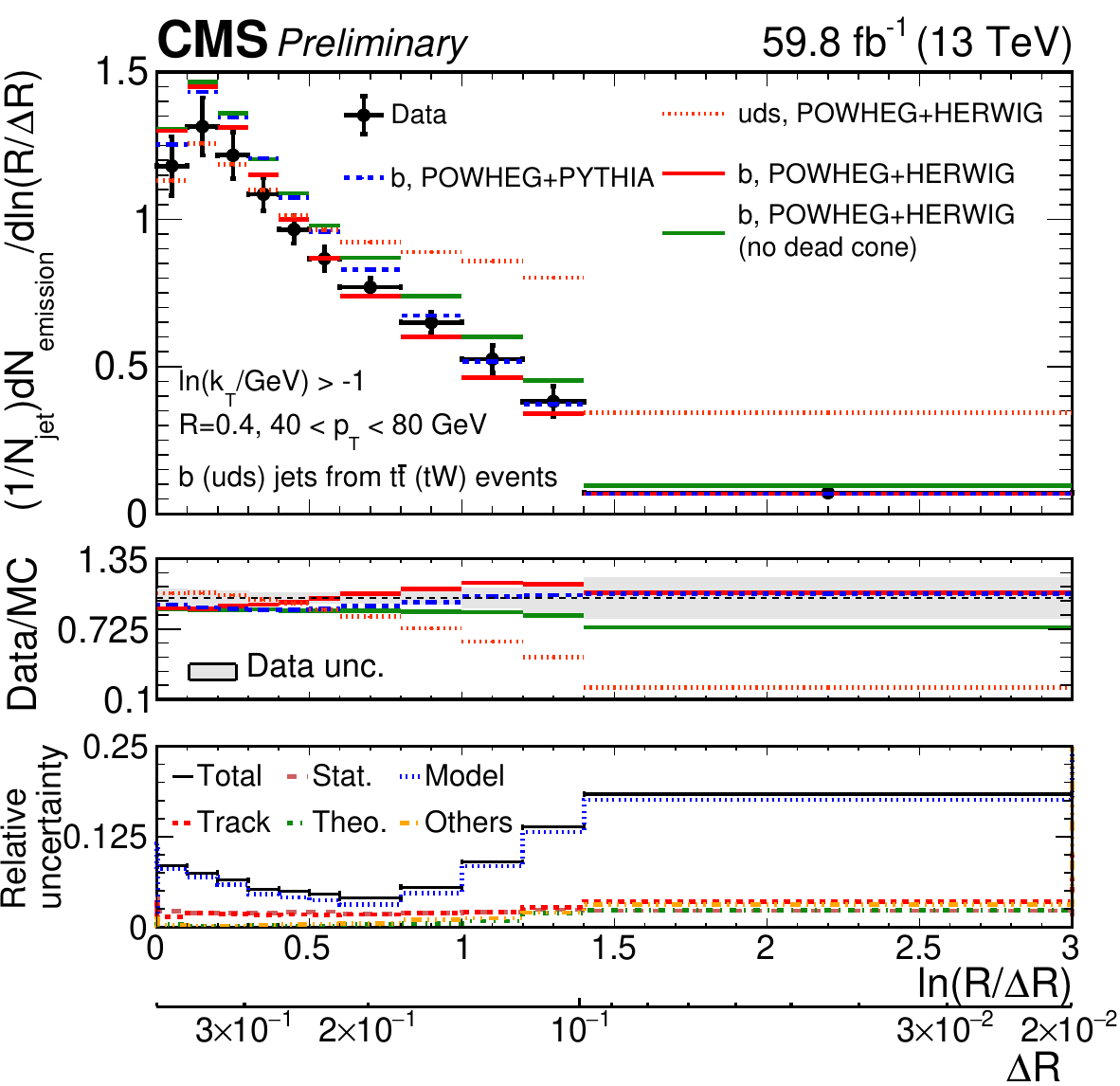}
  \caption{(Top) Ratio of jet cross section calculated from pQCD at NNLO with hadronization correction applied to cross sections at $\sqrt{s}$= 200 GeV and 510 GeV. Plots taken from Ref.~\cite{STAR:InclusiveJets}. (Bottom) The measured emission density is shown differentially in $\ln(R/\Delta R)$ bins for b quark
jets with 40 $<p_{\mathrm{T}}<$ 80 GeV. The vertical error bars rep-
resents the statistical and systematics uncertainties summed in quadrature. The middle panels show the ratio of the data to the MC predictions with a
dashed line at one for reference. The total uncertainties of the data are shown as the gray band. The lower panels show the breakdown of the data uncertainties into various sources. Plot taken from Ref.~\cite{CMS:DeadConePAS}.}
  \label{fig:extensions}
\end{figure}

\begin{figure}[!t]
  \centering
 \includegraphics[width=0.43\linewidth]{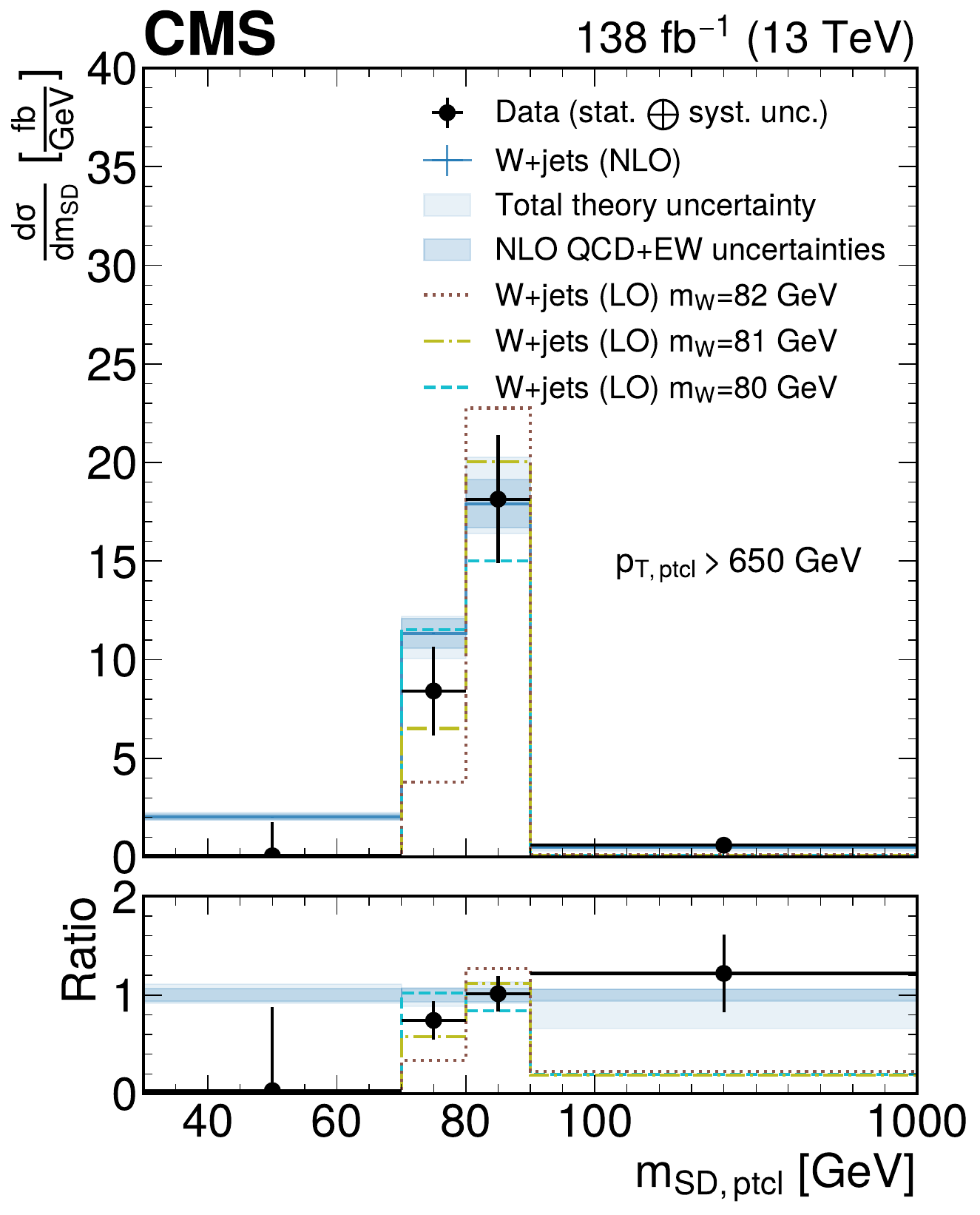}
 \includegraphics[width=0.43\linewidth]{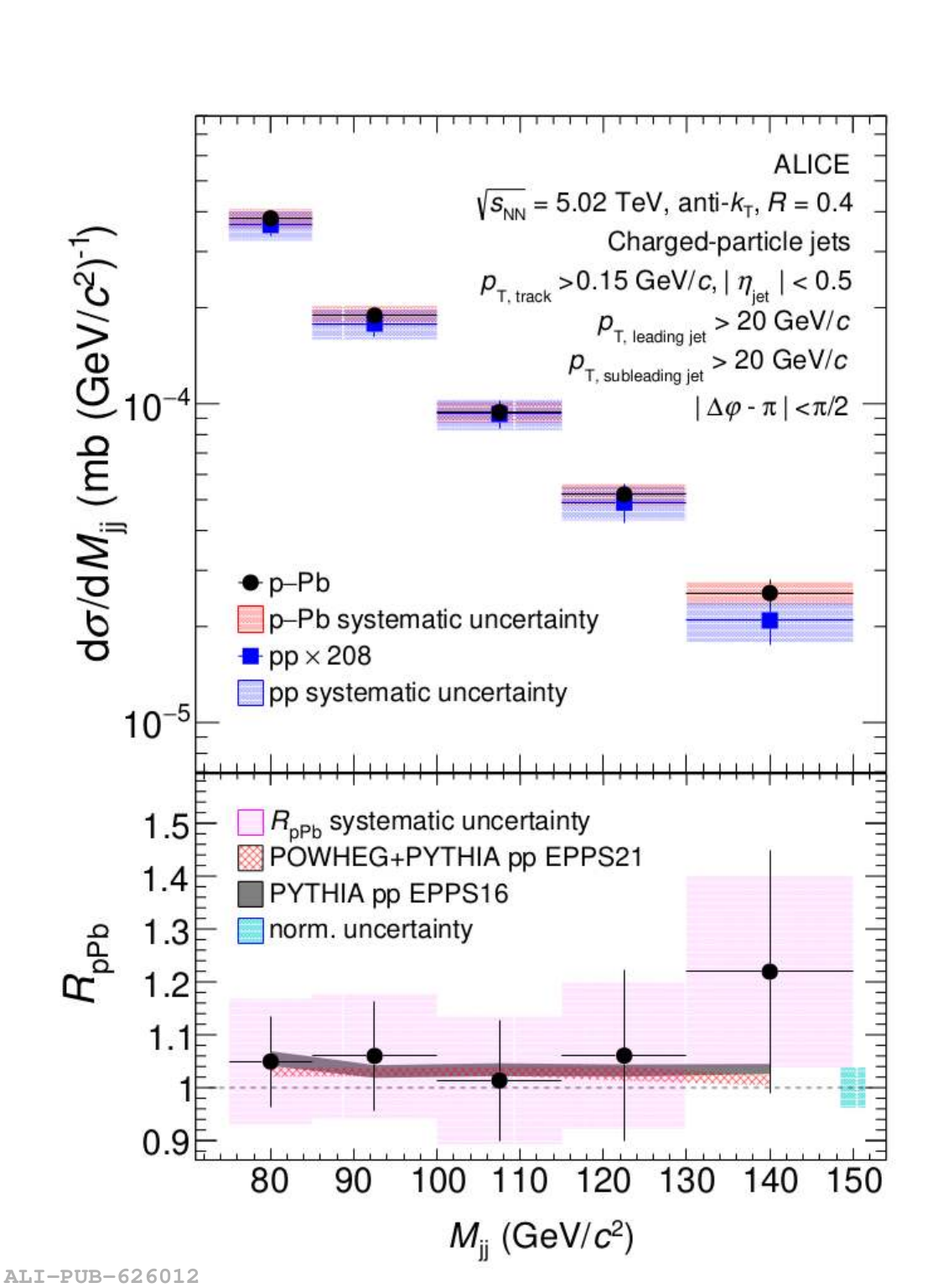}
  \caption{(Left) Unfolded and background subtracted jet mass distribution at the particle level for $p_{\mathrm{T, ptcl}} >$ 650 GeV. The unfolded data are shown as black markers. The theory uncertainty is the sum in quadrature of parton shower variations and the hadronization model uncertainty, as well as the uncertainties in the QCD and EW corrections, and is drawn as a light shaded blue band. The purely perturbative uncertainties are overlaid as a dark, shaded blue band. Predictions with different values of the $W$ boson mass generated with PYTHIA at LO and scaled to match the total number of events in data are overlaid. Plot taken from Ref.~\cite{CMS:BoostedWmass}. (Right) The dijet invariant mass spectrum in pp and p--Pb at $\sqrt{s_{\rm NN}}$ = 5.02 TeV is presented in the top panel, and
the nuclear modification factor $R_{\rm pPb}$ in the bottom panel. In the bottom panel, the simulation results from \texttt{PYTHIA} and \texttt{POWHEG+PYTHIA} are also shown. Plot taken from Ref.~\cite{ALICE:DijetMasspPb}.}
  \label{fig:alice-w}
\end{figure}

\section{Recent extensions of the jet programme}
Jet measurements at lower centre-of-mass energies probe a complementary $x$ range.  STAR reported double-differential inclusive-jet cross sections in $pp$ collisions at $\sqrt{s}=200$ and 510~GeV, with an off-axis cone method used to correct the underlying-event contribution~\cite{STAR:InclusiveJets}.  The measurements extend to $x_T=2p_T/\sqrt{s}$ values where the gluon PDF is less well constrained by TeV-scale hadron colliders.  As shown in Fig.~\ref{fig:extensions} (top), NNLO pQCD predictions with modern PDF sets tend to lie above the data, while HERAPDF gives the closest description among the sets shown; Pythia describes the shape reasonably well but shows normalisation offsets, providing useful input for generator tuning.

Other new results emphasise how jet substructure and flavour extend the high-$p_{\rm T}$ programme.  CMS presented evidence for the dead-cone effect in bottom-quark-initiated jets from top-quark pair events~\cite{CMS:DeadConePAS}.  Fig.~\ref{fig:extensions} (bottom) shows the Lund-plane observable used to expose the depletion of collinear radiation. The comparison to a massless shower sample isolates the perturbative suppression from a heavy quark and gives a 4.3 standard-deviation discrepancy with the massless hypothesis, while the nominal simulation agrees with the data.

Jet substructure also contributes to electroweak precision measurements.  CMS measured the soft-drop mass of boosted hadronic $W$ boson decays in $W$+jets events and extracted $m_W=80.83\pm0.55$~GeV, the most precise all-jets determination at a hadron collider so far~\cite{CMS:BoostedWmass}.  Fig.~\ref{fig:alice-w} (left) shows the unfolded and background-subtracted mass distributions.  The selection isolates a clean resonant region in which templates with different input $m_W$ values bracket the measured spectrum.

At lower jet masses and in small collision systems, ALICE measured the dijet invariant mass of charged-particle jets in pp and p--Pb collisions at $\sqrt{s_{\rm NN}}=5.02$~TeV~\cite{ALICE:DijetMasspPb}.  Fig.~\ref{fig:alice-w} (right) shows the p--Pb and scaled pp spectra together with the nuclear modification factor $R_{\rm pPb}$.  The factor is consistent with unity within uncertainties, in line with previous small-system jet studies.  Nevertheless, the low-mass region is expected to be sensitive to nuclear anti-shadowing effects, so higher-statistics data could make this observable a useful input to future nuclear-PDF constraints.

\section{Outlook}

The present jet programme demonstrates the close connection between experimental precision and theory progress.  Inclusive-jet and dijet data at several centre-of-mass energies are already powerful inputs to PDFs and $\alpha_s$, and NNLO predictions with NP and EW corrections have become essential for their interpretation.  At the same time, the comparison of data to theory is now limited not only by total uncertainties but also by the detailed modelling of their correlations.  Improvements in JES calibration, covariance propagation and public preservation of full correlation information will therefore be critical for future global analyses.

Run~3 will provide substantially larger data samples at high transverse momentum, with total integrated luminosities comparable to or larger than Run~2.  The resulting measurements should sharpen constraints on the high-$x$ gluon, test the consistency of different jet data sets, and improve determinations of $\alpha_s$.  They will also strengthen jet-based studies of boosted electroweak objects, heavy-quark radiation, nuclear modifications and searches for new phenomena.  In this sense, high-$p_{\rm T}$ jet physics remains a central component of the LHC precision and discovery programme.

\end{document}